\documentclass[10pt, conference]{IEEEtran}
\usepackage{cite}
\usepackage{amsmath,amssymb,amsfonts}
\usepackage{graphicx}
\usepackage{textcomp}
\usepackage{xcolor}
\usepackage{verbatim}
\usepackage[utf8]{inputenc}
\usepackage{svg}
\usepackage{newunicodechar}
\newunicodechar{−}{\ensuremath{-}}
\usepackage{tikz}
\usepackage{subfig}
\usepackage{float}
\usepackage{booktabs}
\usepackage{comment}
\usepackage{multirow}
\usepackage{makecell}
\usepackage{threeparttable}
\usepackage{pgfplots }
\usepackage[english]{babel}
\usepackage{graphicx,url}

\usepackage{algorithm}
\usepackage{algorithmic}
\usepackage{mathtools}

\usepackage{makecell}

\def\BibTeX{{\rm B\kern-.05em{\sc i\kern-.025em b}\kern-.08em
    T\kern-.1667em\lower.7ex\hbox{E}\kern-.125emX}}
\usepackage{amsthm}

\theoremstyle{remark}

\ifCLASSINFOpdf
\else
\fi

\makeatletter
\let\ps@IEEEtitlepagestyle\ps@empty
\makeatother

\IEEEoverridecommandlockouts
\begin{document}

\title{Multi-Agent Transformer for Queue-Level XR Traffic Scheduling in TSN Networks}

\author{\IEEEauthorblockN{
Marcos Carvalho\IEEEauthorrefmark{1}\IEEEauthorrefmark{3},
Fatih Temiz\IEEEauthorrefmark{3},
Shavbo Salehi\IEEEauthorrefmark{3}
Melike Erol-Kantarci\IEEEauthorrefmark{3}, Fellow, IEEE,
Daniel F. Macedo\IEEEauthorrefmark{1}
}
\\
\IEEEauthorblockA{\IEEEauthorrefmark{1}Universidade Federal de Minas Gerais, Brazil\\\IEEEauthorrefmark{3}School of Electrical Engineering and Computer Science, University of Ottawa, Ottawa, Canada \\
E-mails:\{marcoscarvalho, damacedo\}@dcc.ufmg.br}, \{ftemi033, ssale038, melike.erolkantarci\}@uottawa.ca\\
}

\maketitle
\begin{abstract}

Time-Sensitive Networking (TSN) and Mobile Edge Computing (MEC) hold strong potential for enabling ultra-reliable low-latency communication for time-sensitive applications, such as eXtended Reality (XR). However, the widespread adoption of XR introduces significant challenges due to co-located services in MEC environments, leading to contention for shared network resources. Moreover, XR traffic types have distinct characteristics and criticality in terms of timing requirements, further increasing the complexity and dynamics of such environments. Although reinforcement learning has shown promise for TSN scheduling optimization in dynamic network scenarios, existing approaches rely on centralized or high-level multi-agent designs and are typically tailored to periodic and predictable industrial traffic, limiting their applicability to XR workloads. As a result, these approaches suffer from (i) limited ability to capture inter-queue dependencies due to coarse-grained control, and (ii) poor adaptability to highly dynamic and heterogeneous XR traffic. To address these gaps, we propose a multi-agent reinforcement learning approach for queue-level XR traffic scheduling. We adopt the multi-agent transformer (MAT) to model inter-queue dependencies via attention over agents’ observations and actions, enabling implicit coordination across heterogeneous co-located XR applications. Our simulation results show that the proposed method outperforms baselines, achieving up to 71.42\% latency reduction and up to 83.2\% reduction in failure rate, while consistently achieving high reliability across all queues.
\end{abstract}


\IEEEpeerreviewmaketitle

\section{Introduction}

In recent years, eXtended Reality (XR) has rapidly expanded across various domains such as gaming, education, and industry~\cite{Temiz2026EdgeLearning}. This expansion poses significant challenges to network infrastructure, as XR applications rely on immersive and continuous sensing with stringent ultra-reliable low-latency communication (URLLC) requirements. Meanwhile, mobile edge computing (MEC) has been a promising approach to support URLLC by reducing latency and improving reliability through proximity to end users~\cite{salehi2024self}. Its integration with time-sensitive
networking (TSN) has been explored to enhance timely service delivery~\cite{carvalho2025performance}.
TSN standards such as IEEE 802.1Qbv~\cite{qbv} further support this by enabling time-aware control of multiple queues in network devices. In practice, MEC environments co-locate multiple XR applications, where each is mapped to distinct TSN queues to enable traffic isolation and fine-grained control. However, this resource-sharing strategy introduces competition among multiple applications for network resources. In addition, the diversity of XR applications creates a highly challenging environment. 
For instance, semantic augmented reality applications (SeAR) impose more stringent deadline constraints, as semantically critical content must be delivered within strict latency bounds, compared to traditional augmented reality (AR) applications, which typically operate under soft real-time (SRT) constraints~\cite{luo2025enhancing}. 
Traditional TSN schedulers based on optimization methods, such as satisfiability modulo theories and integer linear programming, struggle to adapt schedules in a timely manner in such dynamic settings~\cite{stuber2023survey}.\looseness=-1

Alternatively, reinforcement learning (RL) has shown promise in improving TSN scheduling, as it enables adaptive decision-making through interaction with the environment. For example,~\cite{roberty2024configuring} applies proximal policy optimization (PPO) to optimize gate timing across multiple queues, reducing end-to-end latency in dynamic industrial networks. In addition,~\cite{zhou2021mitigation} proposes a DDPG-based approach to adapt time slot durations across the queues, aiming to mitigate transmission failures caused by external factors such as time-synchronization errors. Moreover,~\cite{saruhan2025cooperative} proposes a hierarchical RL approach for jointly managing computational resources and TSN scheduling in edge environments, where a high-level policy allocates resources and a low-level policy controls transmission intervals across the queues. Recent studies explore multi-agent reinforcement learning (MARL) by proposing different levels of agent abstraction to address the complexities of TSN networks. For instance,~\cite{guo2024towards} decouples the TSN scheduling problem into specialized routing and timing agents, enabling real-time traffic scheduling at the device level. Similarly,~\cite{garcia2025multi} implements a multi-agent architecture within a centralized control plane, where routing and scheduling agents collaborate to minimize latency in dynamic environments. Moreover,~\cite{tan2025sharp} models each TSN switch as an autonomous agent responsible for managing local traffic. However, existing designs remain centralized at the device level, limiting their ability to capture complex inter-queue dependencies under shared network resources and heterogeneous XR applications.

To address these limitations, we propose a queue-level MARL framework in which each TSN queue is modeled as an individual agent. The agents learn cooperative scheduling policies by dynamically allocating time-slot durations to minimize deadline misses across heterogeneous XR applications. We leverage a multi-agent transformer (MAT)~\cite{wen2022multi} architecture to model inter-queue dependencies through attention mechanisms, enabling effective coordination among agents. To evaluate the effectiveness of the proposed approach, we compare it against single-agent, MARL, and non-learning-based baselines. The results show that MAT-based scheduling significantly outperforms the baselines by learning a more effective cooperative policy, leading to improved latency reduction up to 71.42\% and up to 83.2\% reduction in failure rate, while consistently maintaining high reliability.

\section{System Model and Problem Formulation}
\label{sec:system_model}

\subsection{System Architecture Overview}
\begin{figure*}[t]
  \centering
  \includegraphics[width=0.9\textwidth]{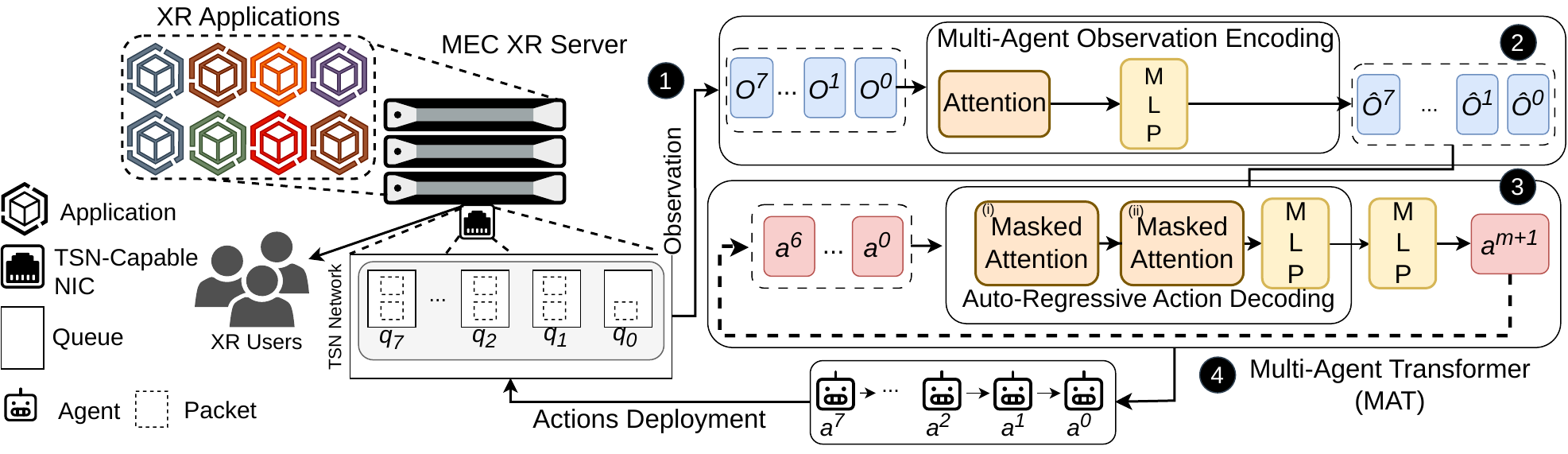}
  \caption{Overall System Model}
  \label{fig:proposedArchitecture}
\end{figure*}

In this paper, we consider an MEC-based XR streaming architecture (Fig.~\ref{fig:proposedArchitecture}) supporting traditional AR and SeAR applications. SeAR applications differ from traditional AR by focusing on the transmission of semantic information rather than raw data~\cite{uysal2022semantic}. Furthermore, we consider different deadline requirements between traditional AR and SeAR applications. We observe that SeAR applications require a more stringent latency bound compared to traditional AR. This observation stems from the fact that semantic frames are typically small but carry an extremely high value of information~\cite{uysal2022semantic,luo2025enhancing}. In addition, we adopt the MEC XR server equipped with a TSN-capable network interface card (NIC)~\cite{carvalho2025performance}. We assume the NIC with a set of queues, denoted by $Q = \{{q_0, q_1, \ldots, q_n}\}$, where each queue serves a specific XR application to ensure strict isolation. Additionally, we consider a downlink (DL) multi-user communication environment in which the number of users for each XR application varies randomly over time.

\subsection{XR Traffic and Latency Model}
XR applications have two temporal variables: inter-frame interval (IFI) and inter-packet interval (IPI). The IFI denotes the time between consecutive video frames and exhibits a non-periodic nature due to external factors~\cite{morin2023extended}. The IPI denotes the time between consecutive packets within a video frame and is modeled as:
\begin{equation}
IPI(p_j^{(i)}) = t_{\text{start}}^{(f)} + \sum_{j=1}^{P-1} \Delta_j + \tau,
\quad P = 1, 2, \dots, N_{\text{IP}}
\label{eq:ipi}
\end{equation}

\noindent where $t_{\text{start}}^{(f)}$ denotes the starting time of video frame $f$, $\Delta_j$ is the interval between consecutive packets within the same frame, $\tau$ captures additional stochastic variations (i.e., IP fragmentation), and $N_{\text{IP}}$ represents the total number of packets in a video frame. These packets are enqueued in the TSN queues according to their IPI and transmitted following a first-in, first-out (FIFO) policy during the transmission opportunity. Hence, latency is mainly determined by the queueing delay, formulated as:
\begin{equation}
W(p_j^{(i)}) = t_s(p_j^{(i)}) - IPI(p_j^{(i)}),
\label{eq:waiting_time}
\end{equation}

\noindent where $t_s(p_j^{(i)})$ denotes the service start time of packet $p_{j}$ in the queue $q_{i}$, and $IPI(p_j^{(i)})$ represents the arrival time of packet $p_j$ in queue $q_i$. Based on the waiting time of all packets in the queue $q_{i}$, the average waiting time is modeled as $\bar{W} = \frac{1}{X_i} \sum_{j=1}^{X_i} W(p_j^{(i)})$, where $X_i$ is the total number of packets. Although $\bar{W}$ reflects the overall latency across all packets, it does not capture worst-case scenarios. The age of the oldest packet (AOP), denoted by $g$, captures this worst-case latency by measuring the older packet in the queues, as formulated by the equation below:
\begin{equation}
g = \max_{p_j \in q_i} W(p_j^{(i)}).
\label{eq:oldest_frame}
\end{equation}
In this work, we consider both IFI and IPI as random variables following a Johnson’s SU distribution~\cite{morin2023extended}. 

\subsection{TSN Schedule Model}
\label{sec:tsn_model}
We discretize time globally into fixed-duration cycles $c_i \in C$, each with duration $C_d = 1$ms. Based on this, a schedule $s_i \in S$ is constructed at each cycle in two steps. First, we define intra-cycle \textit{time-slot allocation} for each queue, where each cycle is partitioned into a sequence of $T$ time slots, with $T \le |Q|$. Formally, we define this step as:

\begin{equation}
T = \big(t_1(q_{0}), t_2(q_{1}), \ldots, t_T(q_{Q})\big), \quad \sum_{j=1}^{T} t_j = C_d,
\label{eq:duration}
\end{equation}

\noindent where $t_j$ denotes the duration of the $j$-th time slot (in $\mu$s) assigned to queue $q_i$. The second step defines the \textit{service order} of queues. We assume that the time-slot duration encodes queue urgency, so queues are scheduled in descending order of $t_j$. We also assume that multiple queues share the same duration, where ties are resolved by prioritizing higher-indexed queues (i.e., $q_{7} > q_{6} > \dots > q_{0}$), following driver-level prioritization~\cite{bosk2022methodology}. The resulting service order is a permutation $\pi(\cdot)$ obtained by sorting queues in decreasing order of $t_k$ and breaking ties by index: $\pi = \operatorname*{arg,sort}_{k \in {1,\dots,Q}} (-t_k, k)$. Thus, the schedule $s_i$ for cycle $c_i$ is defined as:
\begin{equation}
s_i = \big([t_{\pi(1)}, \pi(1)], [t_{\pi(2)}, \pi(2)], \ldots, [t_{\pi(Q)}, \pi(Q)]\big),
\label{eq:schedule}
\end{equation}
\noindent where $t_{\pi(k)}$ is the assigned time-slot duration and $\pi(k)$ denotes the index of the queue scheduled in the $k$-th position.

\subsection{Problem Formulation}
In this work, we assume a hard deadline scenario in which all packets of an XR video frame must be transmitted before the next frame arrives~\cite{paymard2024pdu}. We further assume that SeAR imposes stricter deadlines than traditional applications. Thus, the main objective is to minimize deadline violations in co-located mixed-criticality XR applications by reducing the AOP of each queue relative to its deadline, while a secondary objective is to maximize bandwidth utilization by avoiding allocating more time than is actually needed. The scheduling problem is formulated as follows:

\begin{subequations}\label{eq:objective_TSNScheduling}
\begin{align}
\MoveEqLeft[3] \mathbf{s}^* = \arg\min_{\mathbf{s} \in \mathcal{S}} \sum_{c=1}^{C} 
\Bigg[\sum_{k=0}^{T} \sum_{i=0}^{Q} 
\lambda_1 \frac{g_{i,t_{k}, c}}{D_{q_{i}}} \nonumber \\
& + \sum_{t_k \in c} 
\lambda_2 \big( C(t_{k,q_{i}}) - F_{\text{tx}}(t_{k,q_{i}}) \big) \Bigg] \label{eq:objective_TSNScheduling_obj_main} \\
\text{s.t.} \quad & \sum_{t_k \in c} x_i(t_k) \le 1, \quad \forall i, \forall c, \label{con:one_per_cycle} \\
& F_{\text{tx}}(t_k) \le C(t_k), \quad \forall t_k, \label{con:capacity} \\
& \sum_{t_k \in c} t_k = C_{d}, \quad \forall c, \label{con:cycle_duration} \\
& D_i < IFI_i, \forall i, \label{con:hard_latency}\\
& D_m < D_k, \forall m \in Q_{SeAR}; \forall k \in Q_{AR}, \label{con:sear_ar_deadline}
\end{align}
\end{subequations}

\noindent \noindent where $g_{i,t_{k},c}$ denotes the AOP in queue $q_i$ after completing the transmission of its allocated time slot of duration $t_{k}$ within the cycle $c$. $D_i$ is the deadline of the XR application in $q_i$; For maximizing bandwidth utilization, $C(t_{k,i})$ represents the maximum number of packets that can be transmitted in slot $t_k$ for queue $q_i$, defined as $C(t_k) = \frac{L_{\text{speed}} \cdot t_k}{P_{\text{size}}}$, where $L_{\text{speed}}$ is the link speed, $t_k$ is the slot duration, and $P_{\text{size}}$ is the packet size. $F_{\text{tx}}(t_k)$ denotes the actual number of transmitted frames. 
Eq.~(\ref{con:one_per_cycle}) enforces that each queue is allocated at most once per cycle. Eq.~(\ref{con:capacity}) ensures that the amount of transmitted packets during time slot $t_{k}$ does not exceed the available capacity. Eq.~(\ref{con:cycle_duration}) guarantees that the total duration of all time slots within a cycle equals the predefined cycle duration $C_{d}$. Eq.~(\ref{con:hard_latency}) ensures that the deadline $D_{i}$ of each application does not exceed its IFI. Finally, eq.~(\ref{con:sear_ar_deadline}) enforces stricter deadline requirements for SeAR applications by ensuring that their deadlines are always smaller than those of AR applications.

\subsection{MDP Formulation}
We formulated the TSN scheduling problem as a multi-agent partially observable Markov decision process. Below, we describe the three essential components of an agent.
\begin{itemize}
    \item \textit{Observation:} The local observation $o^i$ of each queue agent $i$ after time slot duration $t_{k}$ is defined as $o^i = \langle F_{tx}, B, g, \bar{W} \rangle$, where $F_{tx}$ is the number of transmitted packets, $B$ is the residual backlog, $g$ is the age of the oldest remaining packet (in $\mu$s), and $\bar{W}$ is the average waiting time of packets (in $\mu$s). 

    \item \textit{Action:} Each queue agent $i$ decides its time-slot allocation by outputting a continuous value $a^i \in [0,1]$ representing its desired fraction of the cycle time. The actual time-slot duration assigned to each queue is then computed as $t_i = \frac{a^i}{\sum_{j=0}^{|Q|} a^j} \cdot C_d$. 

    \item \textit{Reward:} The reward function is designed as a cooperative team reward, encouraging all agents to minimize deadline violations across all queues and maximize the bandwidth utilization. The team reward $R_t$ is formulated as:
\begin{equation}
    R_t = - \max_{i \in \mathcal{Q}} \Phi_i(t) - \beta \frac{1}{Q} \sum_{i=1}^{Q} \mathit{BW}_i(t)
    \label{eq:global_reward}
\end{equation}
\noindent where the term $\Phi_i(t)$ represents the queueing latency penalty, which is computed as follows:
\begin{equation}
    \Phi_i(t) = 
    \begin{cases} 
        \rho_i(t), & \text{if } \rho_i(t) \leq 1 \\ 
        1 + \lambda (\rho_i(t) - 1)^2, & \text{if } \rho_i(t) > 1 
    \end{cases}
    \label{eq:reward}
\end{equation}
where $\rho_i(t)$ is the AOP-deadline ratio in queue $i$, defined as $\rho_i(t) = \frac{g_i(t)}{D_i}$. Then, the team receives a linear penalty based on the maximum ratio across all queues when no deadline is violated ($\rho_i(t) \leq 1$). Once any deadline is violated ($\rho_i(t) > 1$), the penalty increases quadratically, strongly discouraging deadline violations. The term $\lambda$ is a positive scaling factor that ensures a smooth transition between the non-violation and violation cases. The team is further penalized by the average bandwidth utilization term $\mathit{BW}_i(t)$, which is computed as $C(t_{k,q_{i}}) - F_{\text{tx}}(t_{k,q_{i}})$. $\beta$ controls the relative weight of the bandwidth penalty.
\end{itemize}


\section{MAT-Based TSN Scheduling}
\label{sec:proposal}
\subsection{MAT Framework}
To efficiently address the optimization challenge, we employ MAT~\cite{wen2022multi} framework as shown in Fig.~\ref{fig:proposedArchitecture}, which comprises the following blocks: 1) \textit{Multi-Agent Observation Encoding:} this block receives queue observations ($o^0$, $o^1$, \dots, $o^Q$) in an arbitrary order and then extracts information across them,  where each block includes multi-head self-attention and a feedforward neural network. The output of the encoder is a series of latent representations ($\hat{o}^0$, $\hat{o}^1$, \dots, $\hat{o}^Q$) containing both the individual information of each queue $i$ and complex, high-level interactions between them.  2) \textit{Auto-Regressive Action Decoding:} this block employs an auto-regressive approach to generate a sequence of actions. It consists of a sequence of decoding blocks, where each block uses two masked self-attention mechanisms: (i) the first masked self-attention ensures that, when generating the action, attention is computed only with the previous actions. (ii) the second masked attention mechanism computes attention between the current action and the encoder outputs. The output of the final decoder block is a sequence of joint actions for all queues, which is fed to the final MLP to produce the probability distribution of each queue $q_i$'s action, representing its policy.

\subsection{Training and Inference Procedure}
The MAT algorithm (Algo.~\ref{alg:mat_training}) adopts an actor–critic framework in which the encoder and decoder share a latent representation space. During training, the encoder estimates value functions $V_\phi(\hat{o}^i)$ (line 22), encouraging the network to extract informative features from the TSN environment that are critical for scheduling decisions. Coordinated scheduling policies are learned by jointly processing observations and action sequences, where masked self-attention captures both the global network state and causal dependencies among agents (line 25). The model parameters are then updated in a single optimization step by minimizing the combined loss $L_{\text{Encoder}}(\phi) + L_{\text{Decoder}}(\theta)$. For data collection, local observations $o^i$ for each agent are generated after each time slot duration $t_k$ (lines 5–7). The decoder receives these representations and generates joint actions auto-regressively, where each queue selects a time slot duration conditioned on previous decisions (lines 12–14). The environment then executes the actions, determines the service order, transitions to the next state, and returns a team reward $R_t$ (lines 15–18). These transitions are stored in a replay buffer $\mathcal{B}$ for periodic minibatch updates (line 19). Due to space constraints, we omit the loss functions for the $L_{\text{Encoder}}(\phi)$ and $L_{\text{Decoder}}(\theta)$, as well as the definitions of the value function and joint advantage function, and refer the reader to the original MAT paper~\cite{wen2022multi}.

\section{Performance Evaluation}
\label{sec:performance_evaluation}

\subsection{Experimental Settings and Baselines}
We consider a network scenario composed of eight queues supporting five distinct XR applications. Specifically, queues $q_{0}$, $q_{1}$, $q_{3}$, $q_{4}$, and $q_{7}$ handle high-workload traditional AR traffic. In contrast, the remaining queues carry SeAR traffic with different video resolutions, as summarized in Table~\ref{table:parameters}. 
\begin{algorithm}[htb!]
\caption{MAT-Based TSN Scheduling}
\label{alg:mat_training}
\begin{algorithmic}[1]
\STATE \textbf{Initialize:} Encoder $\{\phi_0\}$, Decoder $\{\theta_0\}$, Replay buffer $\mathcal{B}$.
\FOR{each episode $e \in \{1, 2, \dots, N\}$}
    \STATE Reset environment parameters.
    \FOR{each step $s \in \{1,2,\dots,150\}$}
    \FOR{$t_{k} \in T$}
        \STATE Collect $o^i = \langle F_{tx}, B, g, \bar{W} \rangle$ of queue $i$;
        \ENDFOR
        \STATE Form observation vector $o_s = (o^0, \dots, o^Q)$,
        \STATE \textit{// Inference Phase}
        \STATE Encode observations into $\hat{o}^0, \dots, \hat{o}^Q$.
        \STATE Input $(\hat{o}^0, \dots, \hat{o}^Q_)$ to the decoder.
        \FOR{$m = 0, 1, \dots, Q$}
            \STATE Input $a^0, \dots, a^{m}$ and infer $a^{m+1}$.
        \ENDFOR
        \STATE Execute joint action $(a^0, \dots, a^Q)$
        \STATE Determine the service order by eq.~\ref{eq:schedule}.
        \STATE Execute the schedule in the TSN network
        \STATE Obtain the global reward $R_t$ (eq. ~\ref{eq:global_reward}).
        \STATE Store the tuple $(o, a, R_t)$ into $\mathcal{B}$.
        \STATE \textit{// The Training Phase}
        \STATE Sample a random minibatch of $B$ tuples from $\mathcal{B}$.
        \STATE Compute $V_\phi(\hat{o}^0), \dots, V_\phi(\hat{o}^Q)$ from the encoder.
        \STATE Calculate $L_{\text{Encoder}}(\phi)$
        \STATE Compute the joint advantage function $\hat{A}_t$ based on $V_\phi(\hat{o}^0), \dots, V_\phi(\hat{o}^Q)$ with GAE.
        \STATE Input $\hat{o}^0, \dots, \hat{o}^Q$ and $a^0, \dots, a^Q$, generate $\pi_\theta^{0}, \dots, \pi_\theta^{Q}$ at once with the decoder.\;
        \STATE Calculate $L_{\text{Encoder}}(\phi)$ and $L_{\text{Decoder}}(\theta)$
        \STATE Update the encoder and decoder by minimizing $L_{\text{Encoder}}(\phi) + L_{\text{Decoder}}(\theta)$
    \ENDFOR
\ENDFOR
\end{algorithmic}
\end{algorithm}

\noindent In addition, for traditional AR applications, we defined the deadline as 50\% of their IFI. This choice reflects a latency budget distribution strategy, where the end-to-end latency is divided between the MEC TSN scheduling domain and the remaining system components (e.g., 5G network). For SeAR applications, we adopt a stricter deadline of 1 ms, requiring that packets be delivered within a single scheduling cycle, imposing a hard real-time constraint.
\begin{table}[htb!]
\centering
\caption{MEC and MAT parameters/values}
\resizebox{\columnwidth}{!}{%
\setlength{\tabcolsep}{1pt}
\scriptsize
\begin{tabular}{|c|c|c|c|}
\hline

MEC parameter & value & MAT parameter & value \\ \hline

\begin{tabular}[c]{@{}c@{}}\texttt{Network speed}\\ \texttt{Max packet size}\\ \texttt{\# of users per queue}\end{tabular}
&
\begin{tabular}[c]{@{}c@{}}10 Gbps\\ 1428 bytes\\ 1--15\end{tabular}
&
\begin{tabular}[c]{@{}c@{}}\texttt{entropy\_coef}\\ \texttt{gae\_lambda}\\ \texttt{episode\_length}\end{tabular}
&
\begin{tabular}[c]{@{}c@{}}0.01\\ 0.95\\ 150\end{tabular}
\\ \hline

\texttt{AR setting/deadline}
&
\begin{tabular}[c]{@{}c@{}}q0,q3: 3840$\times$1920@90Hz/5.55 ms\\
q1,q4,q7: 3840$\times$1920@72Hz/6.94 ms\end{tabular}
&
\begin{tabular}[c]{@{}c@{}}\texttt{learning\_rate}\\ \texttt{ppo\_epoch}\end{tabular}
&
\begin{tabular}[c]{@{}c@{}}1e-4\\ 5\end{tabular}
\\ \hline

\texttt{SeAR setting/deadline}
&
\begin{tabular}[c]{@{}c@{}}q2: 1920$\times$720@60Hz/1 ms\\
q5: 1280$\times$480@60Hz/1 ms\\
q6: 2560$\times$960@60Hz/1 ms\end{tabular}
&
\begin{tabular}[c]{@{}c@{}}\texttt{n\_heads}\\ \texttt{n\_block}\\ \texttt{n\_embd}\end{tabular}
&
\begin{tabular}[c]{@{}c@{}}1\\ 2\\ 128\end{tabular}
\\ \hline

\end{tabular}%
}
\label{table:parameters}
\end{table}

\noindent The MAT parameters are set according to commonly used configurations in the literature~\cite{zhou2026multi}. For a fair comparison, all baselines share the same hyperparameters, including learning rate, PPO epochs, and entropy coefficient. Each episode consists of 150 steps, with the number of users per queue randomly varying between 1 and 15. We set $\beta = 0.2$ and $\lambda = 4$ in the reward function. All results in the following subsection are reported as the mean and standard error of the mean over three seeds, with the methods’ performance evaluated over 4800 steps.

\textit{Baselines:} We compare MAT against single-agent, MARL, and non-learning-based heuristics, as described below:
\begin{itemize}
    \item \textbf{Single-agent RL:} PPO and advantage actor-critic (A2C)~\cite{mnih2016asynchronous}, representing centralized control.    
    \item \textbf{MARL:} heterogeneous-agent proximal policy optimization (HAPPO)~\cite{kuba2021trust}, where each queue is an agent.
    \item \textbf{Backlog-aware heuristic (BAH):} prioritizes queues with larger backlogs by allocating longer time slots, computed as $t_i = \frac{B_i}{\sum_{j \in Q} B_j} \cdot C_d$,
    where $B_i$ is the backlog of queue $i$ and $C_d$ is the scheduling cycle duration.
    \item \textbf{Deadline-aware heuristic (DAH):} prioritizes queues closer to their deadlines based on the AOP-deadline ratio, allocating time slots as $t_i = \frac{\rho_i}{\sum_{j \in Q} \rho_j} \cdot C_d$, where $\rho_i$ denotes the AOP-deadline ratio of queue $i$.
\end{itemize}

\subsection{Experimental Results}
\begin{figure}[htb!]
	\centering
    \includegraphics[width=0.7\columnwidth]{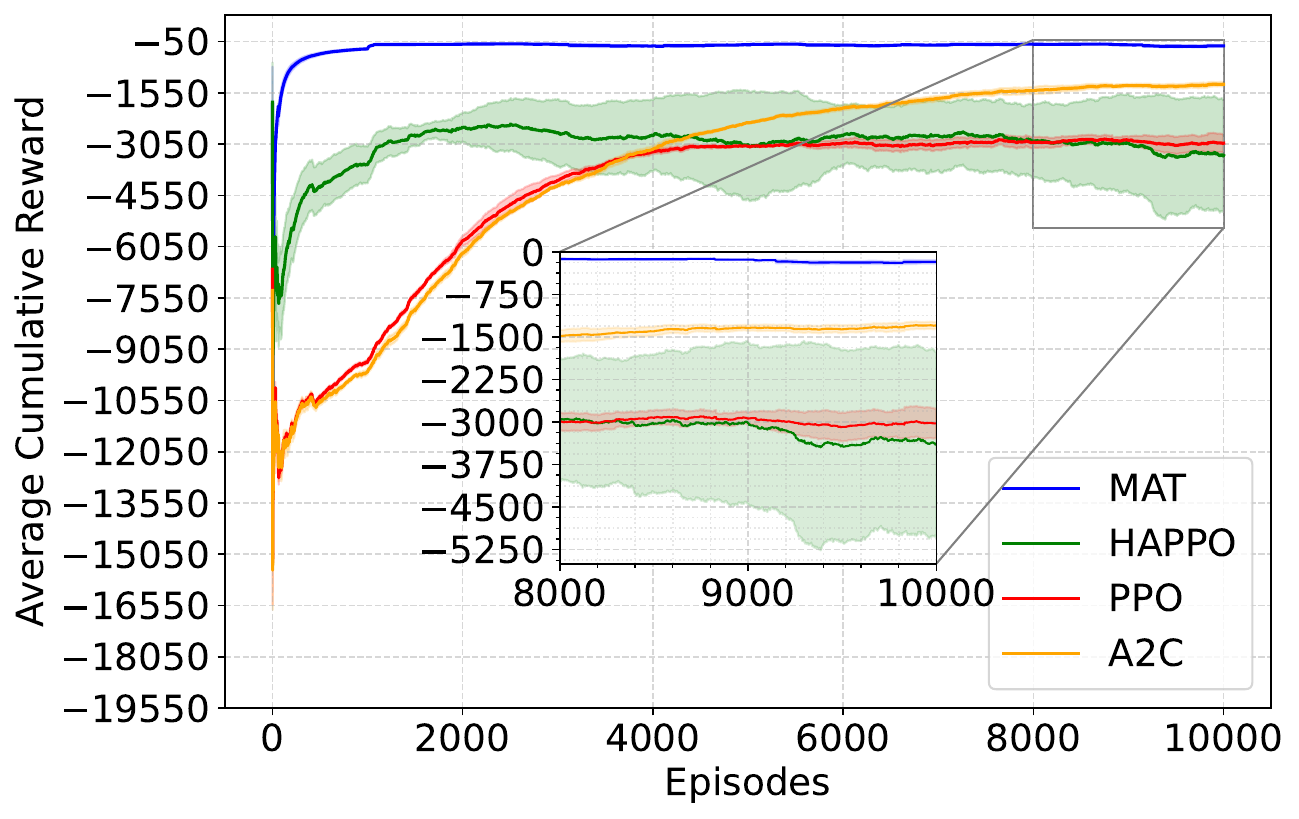}
    \caption{Average Cumulative Reward}    
    \label{fig:reward}
\end{figure}

\textit{Learning performance:} As shown in Fig.~\ref{fig:reward}, MAT achieves higher reward values, faster convergence, and lower variance across different seeds, indicating greater robustness.

\noindent In contrast, HAPPO exhibits higher variance due to its randomly ordered sequential updates, which hinder dependency capture. A2C achieves a significantly higher final reward, which can be explained by the fact that A2C avoids the conservative updates induced by PPO’s clipped objective. However, A2C is limited by its centralized decision-making, which does not explicitly model the interactions among queues. Thus, the superior performance of MAT can be attributed to its sequential decision-making directly into the policy via an auto-regressive formulation, enabling explicit conditioning on previously selected actions and leading to more consistent and globally coordinated behavior.

\begin{figure}[htb!]
	\centering
    \includegraphics[width=1.01\columnwidth]{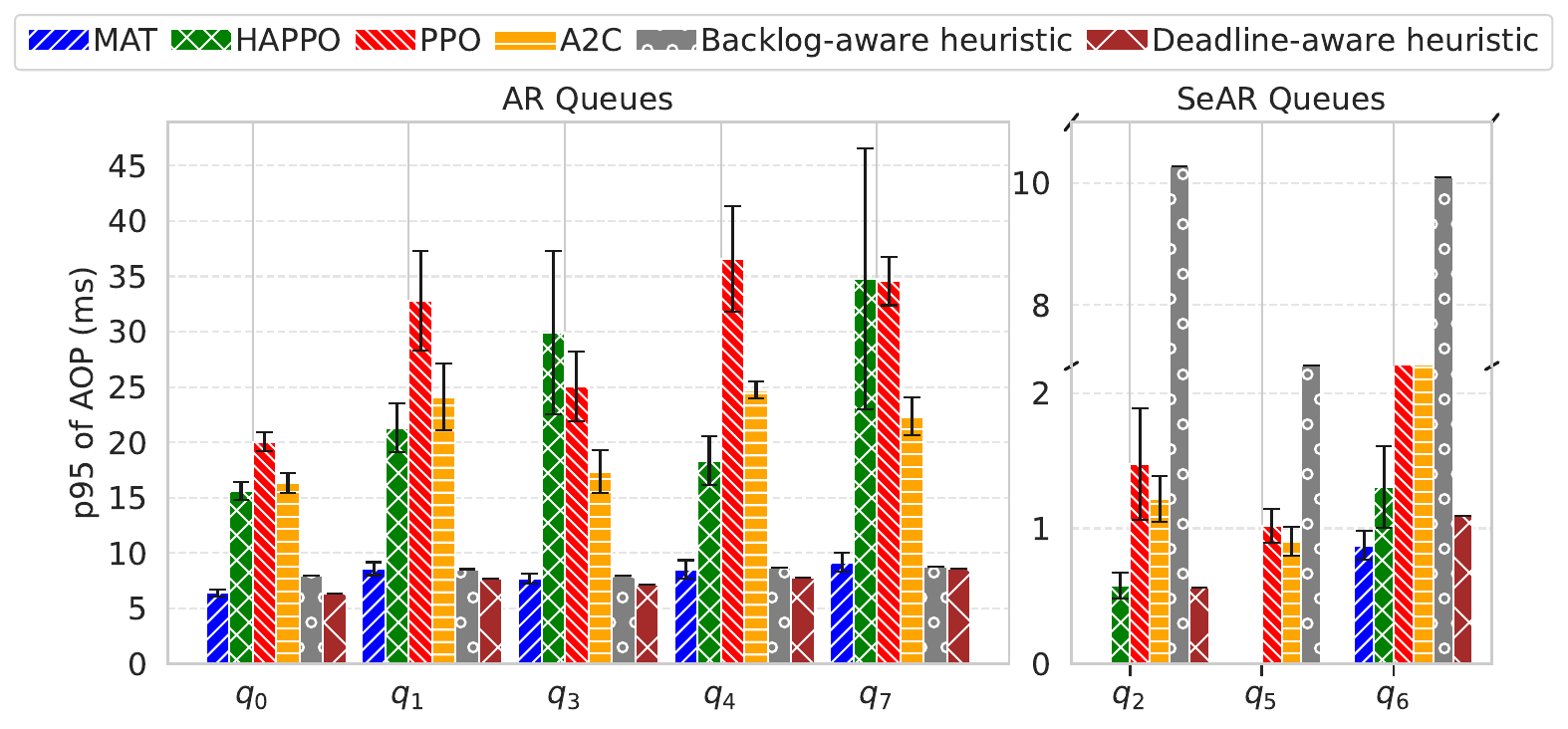}
    \caption{95th percentile of queueing latency}    
    \label{fig:p95}
\end{figure}

\textit{Network-centric performance:} Fig.~\ref{fig:p95} shows the 95th percentile of the AOP across the queues.  
As observed, at the 95th percentile, MAT ensures that packets in a given AR queue are delivered within 10 ms, whereas other RL-based methods range from approximately 15 ms to 35 ms, representing a latency reduction of up to 71.42\%. Moreover, while non-learning baselines match MAT performance on AR queues, they perform poorly overall. As the backlogs on AR queues grow faster, the backlog policy prioritizes them, delaying SeAR packets by up to approximately 10 cycles (10 ms). Conversely, although the AOP-to-deadline ratio increases faster in SeAR queues, the deadline heuristic still incurs delays (up to 0.5 ms in $q_{2}$ and $>$ 1 ms in $q_{6}$), showing that SeAR traffic cannot be effectively prioritized when other queues dominate, forcing both policies to sacrifice SeAR performance. In contrast, MAT coordinates all queues without such trade-offs by allocating time appropriately and responding faster to critical SeAR queues. This highlights the advantage of our queue-level design over centralized high-level single-agent approaches, as well as the ability of attention mechanisms to capture inter-queue dependencies and adapt resource allocation to varying urgency, outperforming HAPPO.

To further assess performance, we evaluate overall service reliability, defined as the percentage of packets meeting their deadlines and expressed in “nines,” as reported in Table~\ref{tab:reliability_1}. 
\begin{table}[htb!]
\centering
\caption{Service reliability (\%) across the queues. Arrows indicate the percentage reduction in failure rate compared to the best baseline method.}
\resizebox{\columnwidth}{!}{%
\begin{tabular}{@{}clccccc@{}}
\toprule
   & MAT & HAPPO & PPO & A2C & BAH& DAH \\ \midrule
$q_{0}$ & \textbf{99.7284}(0.01) {\scriptsize $\uparrow$8.8\%} & 99.3531(0.03) & 99.4832(0.009) & 99.4450(0.01) & 99.6291 & 99.7021  \\
$q_{1}$ & 99.6697(0.009) {\scriptsize $\downarrow$3.8\%} & 99.3110(0.02) & 99.4540(0.01) & 99.4072(0.01) & 99.5472 & \textbf{99.6723}  \\
$q_{2}$ & \textbf{99.9692}(0.02) {\scriptsize $\uparrow$83.2\%} & 99.6603(0.02) & 99.7302(0.02) & 99.5885(0.03) & 97.2819 & 99.7719  \\
$q_{3}$ & \textbf{99.7238}(0.006) {\scriptsize $\uparrow$8.3\%} & 99.3258(0.03) & 99.4900(0.01) & 99.4407(0.02) & 99.6829 & 99.6986  \\
$q_{4}$ & \textbf{99.6699}(0.006) {\scriptsize $\uparrow$4.3\%} & 99.2570(0.02) & 99.4593(0.01) & 99.4244(0.01) & 99.5527 & 99.6551  \\
$q_{5}$ & \textbf{99.9527}(0.009) {\scriptsize $\uparrow$61.4\%} & 99.8838(0.03) & 99.6782(0.01) & 99.6446(0.01) & 97.0370 & 99.8773  \\
$q_{6}$ & \textbf{99.8564}(0.02) {\scriptsize $\uparrow$39.8\%} & 99.4185(0.05) & 99.6206(0.02) & 99.5222(0.01) & 98.9155 & 99.7559  \\
$q_{7}$ & \textbf{99.7026}(0.007) {\scriptsize $\uparrow$3.2\%} & 99.2064(0.07) & 99.4591(0.01) & 99.4142(0.01) & 99.6343 & 99.6929  \\ \bottomrule
\end{tabular}%
}
\label{tab:reliability_1}
\end{table}

\begin{figure}[htb!]
	\centering
    \includegraphics[width=1\columnwidth]{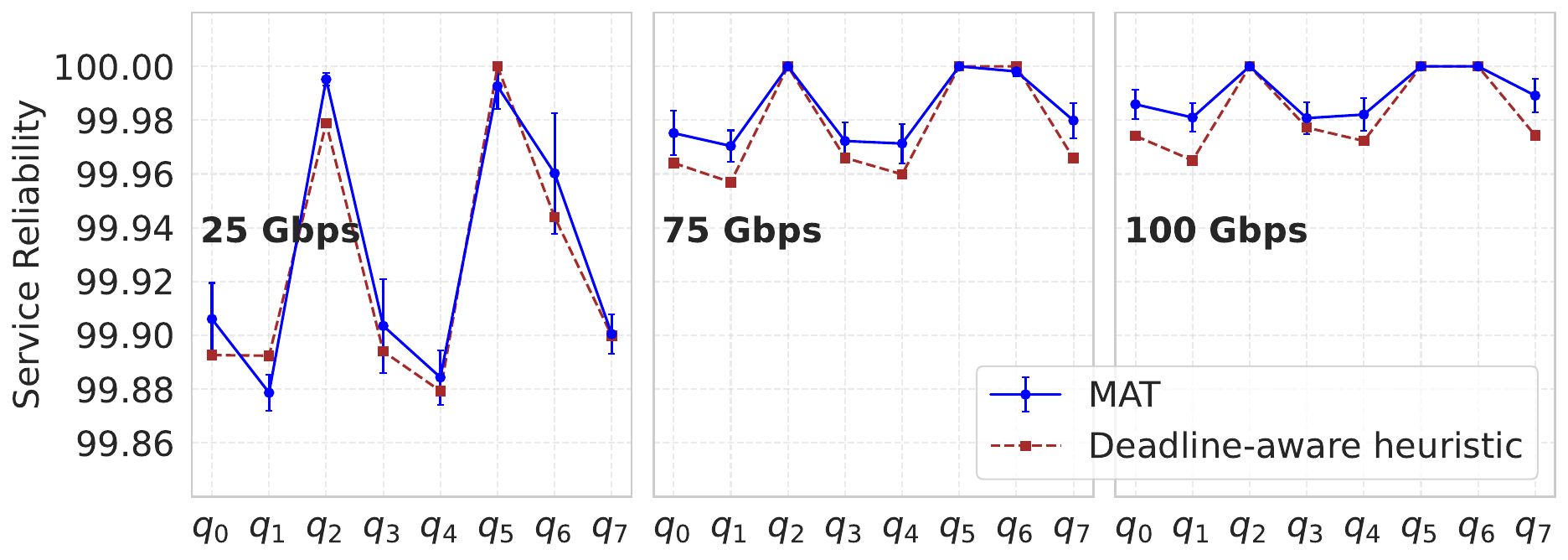}
    \caption{Service reliability (\%) under different network speed}    
    \label{fig:reliability_2}
\end{figure}

\noindent As shown, MAT consistently outperforms all baselines, achieving reliability improvements of up to 0.2–0.4 percentage points. This improvement translates into a reduction in failure rates ranging at least from 3\% to 8\%. Larger performance gaps are observed in the SeAR queues, where the difference between MAT and the baselines can reach a shift from two-nines to three-nines reliability, representing a reduction in failure rate of up to 83.2\%. This improvement can be attributed to the superior cooperative policy learned by MAT, driven by the reward design, which enforces coordination among queues and encourages the agent to allocate resources according to their relative urgency. Due to high contention, 100\% reliability is not expected. We thus evaluate MAT and DAH under increased network speeds to assess performance under higher capacity, as shown in Fig.~\ref{fig:reliability_2}. As expected, reliability increases for both methods as network speed increases, since more packets can be transmitted per cycle. However, the results indicate that the deadline-aware heuristic requires higher network capacity to achieve reliability levels that are already attained by MAT under lower bandwidth conditions, suggesting better efficiency and scalability of the proposed approach. Additionally, MAT demonstrates stronger generalization across different network speeds, maintaining consistently high reliability.

Finally, we evaluate the MAT inference time on an RTX 5090 GPU (107.6 TFLOPS, 31.8 GB VRAM) and an RTX PRO 6000 S GPU (93.6 TFLOPS, 95.6 GB VRAM). On the RTX 5090, MAT achieves a latency of 1.26 ± 0.037 ms, while on the RTX PRO 6000 S it achieves 0.9777 ± 0.0039 ms. Indeed, MAT needs high-performance hardware to keep inference latency below the cycle time, mainly due to sequential autoregressive decoding across queues and transformer overhead.

\section{Conclusion}
This paper addresses scheduling challenges arising from the co-location of heterogeneous XR applications in MEC-TSN integration. We propose a queue-level MARL framework that models inter-queue dependencies using attention and auto-regressive decoding, enabling effective cooperative TSN scheduling. Simulation results show that MAT outperforms baselines, reducing queueing latency and achieving satisfactory reliability. Unlike greedy policies that sacrifice strict SeAR performance, MAT effectively balances heterogeneous deadlines, highlighting the potential of transformer-based MARL for URLLC in XR applications. Future work may investigate advanced knowledge and policy distillation techniques to further reduce inference latency.

\section*{Acknowledgment}
This work was financed in part by the Coordenação de Aperfeiçoamento de Pessoal de Nível Superior - Brasil (CAPES) - Finance Code 001, CNPq (funding agency from the Brazilian federal government), FAPEMIG (Minas Gerais State Funding Agency), and São Paulo Research Foundation (FAPESP) with Brazilian Internet Steering Committee (CGI.br), grants 2018/23097-3 and 2020/05182-3, NSERC Canada
Research Chairs Program. The authors also gratefully acknowledge the support of the University of Ottawa Visiting Researchers Program/Office.


\bibliographystyle{IEEEtran}
\bibliography{bibtex}

\end{document}